\def\o{\over}
\def\Ar{\rightarrow}
\def\bar{\overline}

\def\a{\alpha}
\def\b{\beta}
\def\n{\nu}
\def\m{\mu}

\def\th{\theta}

\def\bar{\overline}

\def\G{{\rm GeV}}

\def\eV{{\rm eV}}
\documentstyle [12pt,epsf]{article}
\begin{document}
\baselineskip=24.5pt
\setcounter{page}{1}
\thispagestyle{empty}
\topskip 2.5  cm
%\topskip 0.5  cm
%\begin{flushright}
%\begin{tabular}{c c}
%& {\normalsize hep-ph/9906516  EHU-99-6}\\
%& June 1999
%\end{tabular}
%\end{flushright}
\vspace{1 cm}
\centerline{\Large\bf  Search for CP Violation with a Neutrino Factory}
\vskip 1.5 cm
\centerline{{\bf Morimitsu TANIMOTO}
  \footnote{E-mail address: tanimoto@edserv.ed.ehime-u.ac.jp}}
\vskip 0.8 cm
 \centerline{ \it{Science Education Laboratory, Ehime University, 
 790-8577 Matsuyama, JAPAN}}
\vskip 4 cm
\centerline{\bf ABSTRACT}\par
\vskip 0.5 cm
 We have discussed the search of the genuine $CP$ violation 
 in a neutrino factory, in which 
well known neutrino beams are provided by a high intensity muon storage ring. 
   Both $\Delta P(\n_\a \Ar \n_\b)$ and  $CP$ odd asymmetry $A_{CP}$
 in neutrino oscillations 
   are investigated  by taking account of the atmospheric neutrino data and 
 the solar neutrino data.
  If  the large mixing angle MSW  solution is taken,  
    the magnitude of $\Delta P(\n_e \Ar \n_\m)$ could be  $1\%$.
        If $s_{13}$ is lower than $0.05$ with the maximal $CP$ violating
 phase,  the matter effect is negligible  in $\Delta P(\n_e \Ar \n_\m)$
 and $A_{CP}$.
        We have proposed how to extract the genuine $CP$ violation effect
    in  the three family scheme 
         from the  neutrino oscillation data $\Delta P(\n_e \Ar \n_\m)$,
     $\Delta P(\n_\m \Ar \n_e)$  and $\Delta P(\n_e \Ar \n_\tau)$.
         
\newpage
\topskip 0  cm
%%%%%%%%%%%%%%%%%%%%%%%%%%%%%%%%%%%%%%%%%%%%%%%%%%%%%%%%%%%%%%%%%%%%%%%%%%%%%%%
%%%%%%%%%%%%%%%%%%%%%%%%%%%%%%%%%%%%%%%%%%%%%%%%%%%%%%%%%%%%%%%%%%%%%%%%%%%%%%%

  Neutrino flavor oscillations provide 
   information of the fundamental property of neutrinos such as masses, flavor
 mixings  and  the $CP$ violating phase. Recent experimental data of neutrinos
  make big impact on  these property.
   Most exciting one is the results at Super-Kamiokande on the  atmospheric 
 neutrinos,  which indicate the large neutrino flavor oscillation of 
 $\n_\m\Ar \n_x$ \cite{SKam}.
   Solar neutrino data also provide the evidence of the neutrino oscillation, 
 however this problem is still uncertain \cite{BKS}. 
 Now, a new stage is represented by the long baseline (LBL) neutrino 
oscillation   experiments.
   The LBL accelerator experiment K2K \cite{K2K} begins taking data
 in this year (1999), whereas the MINOS \cite{MINOS} 
 and a CERN to Gran Sasso project \cite{ICARUS} will start in the first year 
 of the next century.
 
   Some authors \cite{MT,AS,MiNu,Bi,Barger} have already discussed  
   possibilities of observing $CP$ violation in  LBL experiments by 
 measuring  the difference of transition probabilities between $CP$-conjugate
    channels \cite{CP0,CP},  
 which originates from the phase of the neutrino mixing matrix \cite{MNS},
         such as $\n_\m \Ar\n_e$ and $\bar \n_\m \Ar \bar \n_e$.
          However, the direct measurement is very difficult in the planned LBL
  experiments since the magnitude of its difference is usually expected 
 below $0.01$ and the difference of
          energy distributions of neutrino beams $\n_\m$ and
 $\bar\n_\m$  disturbs this measurement in the order of  ${\cal O}(0.01)$.
 Moreover, the matter effect due to the earth makes difficult to extract
 the genuine $CP$ violating effect from the neutrino oscillation data. \par
 
  On the other hand, {\it a neutrino factory} \cite{Geer,Rujula}, in which 
   excellent neutrino beams are provided  by a high intensity muon storage 
 ring, is planned.  These experiments may make possible to search for  $CP$ 
 violation
   in  neutrino oscillations because  neutrinos from muon decays are well known
   as to flavors and their energy distributions when the muon polarization is 
 known \cite{Rujula}.
 For instance, the averaged energy of $\bar\n_e$  is $7\ (12)\G$
in the decay  of the muon with  $10\ (20)\G$.
Based on this set up with $L=732{\rm Km}$, 
the observability of $CP$ violation was
studied in some  mixing angles in ref.\cite{Rujula}.

  In this letter, we estimate the magnitude of $CP$ violation
 in the neutrino factory by using  relevant parameters, in which the $CP$
violating effect is sizable, and then   discuss how to extract the 
 genuine $CP$ violation
         from  neutrino oscillation data, which include the matter effect.
  In order to estimate the $CP$ violating effect in the standard model, 
 we consider  the three family model without sterile neutrinos.
         Our starting point as to the neutrino mixing is
  the large $\nu_\m \Ar \nu_\tau$ oscillation of  atmospheric neutrinos  with 
 $\Delta m^2_{\rm atm}=  (2\sim 6)\times  10^{-3} \eV^2$ and 
 $\sin^2 2\th_{\rm atm} \geq 0.84$,
 which are derived from the recent data of the atmospheric neutrino deficit 
 at Super-Kamiokande
   \cite{SKam}. 
   In the solar neutrino problem \cite{BKS},
 we consider three solutions:
  the small mixing angle MSW solution, the large mixing angle MSW  solution 
 and the vacuum oscillation solution.
 These mass difference scales are 
 $\Delta m_{\odot}^2= 10^{-10}\sim 10^{-4}\eV^2$,
  which are  much smaller than the atmospheric one.
  We put $\Delta m_{\rm atm}^2=\Delta m_{32}^2$ and 
  $\Delta m_{\odot}^2=\Delta m_{21}^2$,
   and so disregard the LSND data \cite{LSND}.
  \par

 %%%%%%%%%%%%%%%%%%%%%%%%%%%%%%%%%%%%%%%%%%%%%%
 %%%%%%%%%%%%%%%%%%%%%%%%%%%%%%%%%%%%%%%%%%%%%%
  If neutrinos are  Majorana particles,  one finds three $CP$ violating phases.
    However, the effect of extra Majorana phases is suppressed
           by the factor $(m_\n/E)^2$ \cite{MJ}.
Therefore, $CP$ violation in the neutrino flavor oscillations relates directly 
        to the $CP$ violating phase $\phi$ in the following mixing matrix $U$
   for  massive neutrinos:
 
  \begin{equation}  
  U = \left (\matrix{ c_{13} c_{12} & c_{13} s_{12} &  s_{13} e^{-i \phi}\cr 
  -c_{23}s_{12}-s_{23}s_{13}c_{12}e^{i \phi} & 
 c_{23}c_{12}-s_{23}s_{13}s_{12}e^{i \phi} &     s_{23}c_{13} \cr
  s_{23}s_{12}-c_{23}s_{13}c_{12}e^{i \phi} &
 -s_{23}c_{12}-c_{23}s_{13}s_{12}e^{i \phi} & c_{23}c_{13} \cr} \right ) \ ,
        \label{Mix}
\end{equation} 

\noindent   where  $s_{ij}\equiv \sin{\theta_{ij}}$ and $c_{ij}\equiv 
\cos{\theta_{ij}}$ are vacuum mixings. 
   The amplitude of $\n_\a\Ar \n_\b$ transition with the neutrino  energy $E$
   after traversing the distance $L$ can be written as
  \begin{equation}        
 {\cal{A}}(\n_\a\Ar \n_\b) = e^{-i EL} \left\{\delta_{\a\b} 
   + \sum_{k=2}^3  
 U_{\a k} U_{\b k}^* \left [\exp{\left (-i {\Delta m^2_{k1} L \o 2 E}\right )}
           -1 \right ] \right\} \ , 
                 \label{Pro}
\end{equation}
\noindent where $U_{\a i}$ is a neutrino mixing matrix element 
 in eq.(\ref{Mix}) \cite{MNS}, in which $\a$ and $i$  refer to the flavor 
 eigenstate and the mass eigenstate, respectively. 
  The  amplitude  ${\cal A} (\bar\n_\a\Ar \bar\n_\b)$ is given by 
  replacing $U$ with $U^*$ in the right hand side in eq.(\ref{Pro}).
 Direct measurements of $CP$ violation originated from the phase $\phi$ 
 are  differences of  transition probabilities between $CP$-conjugate
    channels \cite{CP0,CP}:  
%%%%%%%%%%%%%%%%%%%%%%%%%%%%%%%%%%%%%%%%%%%%%%%%%%%%%% 
  \begin{eqnarray}        
  \Delta P_{CP} &\equiv& P(\bar \n_\m \Ar \bar \n_e) - P(\n_\m \Ar \n_e) =
    P(\n_\m \Ar \n_\tau) -  P(\bar \n_\m \Ar \bar \n_\tau)  \nonumber \\
 &=& P(\bar \n_e \Ar \bar \n_\tau)- P(\n_e \Ar \n_\tau)    
   =  4 J^{\n}_{CP} f_{CP} \ ,
   \label{CP}
   \end{eqnarray}
  \noindent
   where   $f_{CP}$ is the sum of  oscillatory terms as
   \begin{equation}
 f_{CP}\equiv \sin \Delta_{12}+\sin \Delta_{23}+\sin \Delta_{31} \ ,
 \label{SCP}
   \end{equation}
   \noindent with 
  \begin{equation}
 \Delta_{ij}=  \Delta m^2_{ij}{L\o 2E} \ ,
 \label{D}
   \end{equation}
 \noindent
 and the  rephasing  invariant quantity $J^{\n}_{CP}$ is given as \cite{J}
         \begin{equation}
    J_{CP}^{\n}= s_{12} s_{23} s_{13} c_{12} c_{23} c_{13}^2 \sin\phi \ .
        \end{equation}
   \noindent  Since  oscillatory terms are periodic in $L/E$ and  
   $\Delta_{12}+\Delta_{23}+ \Delta_{31}=0$ is satisfied,
   these terms tend toward cancellation among them.
    \par
        
%%%%%%%%%%%%%%%%%%%%%%%%%%%%%%%%
How large is $J^{\n}_{CP} f_{CP}$?
The magnitude $f_{CP}$ depends on $\Delta m_{\rm atm}^2$ and 
$\Delta m_{\odot}^2$.
The detail behavior was discussed in ref.\cite{AS}.
We show the neutrino energy dependence of  $f_{CP}$ in the case
 $\Delta m_{\rm atm}^2=3\times 10^{-3}, \ 6\times 10^{-3}\eV^2$ and 
$\Delta m_{\odot}^2=10^{-5},  \ 10^{-4}\eV^2$ 
 with $L=732 {\rm Km}$ in fig.1. The expected magnitude of  $|f_{CP}|$  is
  at most $0.06$ in $E\geq 5\G$. 
The magnitude $f_{CP}$ is suppressed as $\Delta m_{\rm atm}^2$ and 
$\Delta m_{\odot}^2$  decrease.

   On the other hand, $J^{\n}_{CP}$ depends on  mixing angles.
   The mixings $s_{23}$,  $s_{12}$ and  $s_{13}$ 
   are constrained by  atmospheric neutrinos, solar neutrinos and
   CHOOZ experiments \cite{CHOOZ}, respectively.
       We take $s_{23}=1/\sqrt{2}$, which is the typical mixing angle 
from the atmospheric neutrino data, and $s_{13}\leq 0.2$  from the CHOOZ data.
  Since larger  mixings  give  larger $J^{\n}_{CP}$, 
the large mixing angle solution of the
  solar neutrino is favored to search for  $CP$ violation.
  Putting $s_{12}\simeq 0.5$, which is a typical 
 large mixing angle MSW  solution,  we get  $|J^{\n}_{CP}|\leq 0.04 \sin \phi$.

 The magnitude of  $\Delta P$ can reach  $10^{-3}\sim 10^{-2}$  
in the case of   $\Delta m_{\odot}^2\simeq 10^{-4}\eV^2$  \cite{AS}. 
  The vacuum oscillation solution is unfavor for observing $CP$ violation 
 since  $f_{CP}$ is considerably suppressed
   due to $\Delta m^2_{\odot}\simeq 10^{-10}\eV^2$.
  Then, $\Delta P$ is at most $10^{-8}$. 
   If we use the small angle MSW solution,  $\Delta P$ is at most $10^{-4}$
due to the small mixing angle $s_{12}\sim 0.04$.  
 In conclusion, one can expect  to observe the $CP$ violating effect
in  neutrino oscillations  when the large mixing angle MSW solution is the
true solution in the solar neutrino problem. 
 Therefore, we show numerical calculations by taking this solution
in this letter.
   
 In practice,  a realsitic observable may be the  $CP$ odd asymmetry $A_{CP}$
   \cite{CP0,Rujula}:
  \begin{equation}
  A_{CP}\equiv \frac{P(\n_\a \Ar \n_\b) - P(\bar\n_\a \Ar \bar\n_\b)}
                    {P(\n_\a \Ar \n_\b) + P(\bar\n_\a \Ar \bar\n_\b)} \ .
 \end{equation}
\noindent
 Even if $\Delta P_{\rm CP}$ is smaller than  $10^{-3}$,
the asymmetry could be ${\cal O}(1)$. We also calculate  $A_{CP}$ as well as
$\Delta P_{\rm CP}$ in this letter.
 
 %%%%%%%%%%%%%%%%%%%%%%%%%%%%%%%%%%%%%%%%%%%%%%%%%%%%%%%%%%%
 %%%%%%%%%%%%%%%%%%%  Matter Effect  %%%%%%%%%%%%%%%%%%%%%%%
 %%%%%%%%%%%%%%%%%%%%%%%%%%%%%%%%%%%%%%%%%%%%%%%%%%%%%%%%%%%
 
 Even if  the distance travelled by neutrinos is less than $1000 {\rm Km}$ in
  LBL experiments, those data  include  the background
 matter effect which is not $CP$ invariant.  
 The matter effect should be carefully analyzed since the effect  
  depends strongly on the mass hierarchy, mixings and
 the incident energy of the neutrino as shown in previous works 
 \cite{matter,matter1,matter2}.
 The effective mass squared in the matter $M_m^2$ for
 the neutrino energy $E$ in weak basis is

 %%%%%%%%%%%%%%%%%%%%%%%%%
\begin{equation}  
 {\bf M_m^2} = U \left (\matrix{ m_1^2 & 0 & 0 \cr 
            0 & m_2^2 & 0 \cr
            0 & 0 & m_3^2 \cr} \right )U^\dagger +
      \left (\matrix{ a & 0 & 0 \cr 
            0 & 0 & 0 \cr
            0 & 0 & 0 \cr} \right )  \ , 
                        \label{mass}
\end{equation}

\noindent   
where $a\equiv 2\sqrt{2} G_F n_e E$.
 For antineutrinos, the effective mass squared is given by
 replacing $a\Ar -a$ and $U \Ar U^*$. 
%%%%%%%%%%%%%%%%%%%%%%%%%%%%%%%%%%%%%%%%%%%%%%%%%%%%%%%%%%%%%%%%%%%%%%
 Taking  the constant  matter density $\rho=2.8 g/cm^3$,
   the effective mixing angles and the phase are given 
 in terms of vacuum mixings and the effective neutrino masses,
  which are eigenvalues in eq.(\ref{mass}) \cite{matter2}.
%%%%%%%%%%%%%%%%%%%%%%%%%%%%%%%%%%%%%%%%%%%%%%%%%%%%%%%%%%%%%%%%%%%%%%
It may be important  to note that the constant matter density, assumed in
our calculations, is not always quantitatively accurate
and that a real earth model must eventually be used in the calculation of 
the matter effects.
Actually, Koike and Sato have  discussed  the matter effect
in the K2K experiment  by using the  real earth model \cite{KoSa}.
%%%%%%%%%%%%%%%%%%%%%%%%%%%%%%%%%%%%%%%%%%%%%%%%%%%%%%%%%%%%%%%%%%%%%%
 
 In the neutrino factory, one has a $\n_\m + \bar\n_e\ (\bar\n_\m + \n_e)$ beam
 in the decay of $\m^- \ (\m^+)$. 
 The search for the $CP$ violation effect is  possible
 in different four  oscillation channels. Although the magnitude of
 the genuine $CP$ violation
is expected to be same in all channels as seen in eq.(\ref{CP}),
 matter effects are different in general.
The matter effect may enhance 
$P(\n_\a \Ar \n_\b) - P(\bar\n_\a \Ar \bar\n_\b)$ in the one channel,
but suppress it in the other one.
Therefore, it is important to search for 
$P(\n_\a \Ar \n_\b) - P(\bar\n_\a \Ar \bar\n_\b)$ in different  channels.

%%%%%%%%%%%%%%%%%%%%%%%%%%%%%%%%%%%%%%%%%%%%%%%%%%%%%%%%%%%%%%%%%%%%%%
At first,  we show the result of $\Delta P(\n_e \Ar \n_\m)$ in the matter 
as well as  in vacuum in fig.2 by taking reference set-up  
 $L=732 {\rm Km}$ and $E=7 \G$, which leads to  $a=1.5\times 10^{-3}\eV^2$.
   The large mixing angle MSW  solution is taken for relevant parameters,
     $s_{12}=0.5$, $s_{23}=1/\sqrt{2}$ and 
   $\phi=-90^{\circ}$, $\Delta m_{32}^2 =3\times 10^{-3}\eV^2$ and 
    $\Delta m_{21}^2 =10^{-4}\eV^2$.
 It is remarked that the matter effect increases as $s_{13}$ increases 
 and the magnitude of $\Delta P(\n_e \Ar \n_\m)$ is at most $0.25\%$
as seen in fig 2. However, it could reach to $1\%$
if we take  $\Delta m_{32}^2 =6\times 10^{-3}\eV^2$,
 which is the upper bound of the experimental data.
        The matter effect becomes negligible
        if $s_{13}$ is lower than $0.05$ with $\phi=-90^{\circ}$. 
        
        On the other hand, the $CP$ odd asymmetry $A_{CP}$ of the same process
         is large as seen in fig.3, in which
        parameters are taken to be same ones as in fig.2.
        The matter effect is unimportant  if  $s_{13} \leq 0.05$.
    The larger $A_{CP}$ corresponds to  the smaller absolute value of 
 the oscillation probability.  
 We show these absolute values as well as the $A_{CP}$ 
        in the matter and vacuum in table 1. 
 If $A_{CP}$ is larger than $20\%$, the magnitude  of 
        $P(\n_e \Ar \n_\m)$  is  smaller than $10^{-3}$, which means that 
  the observation of signals of  $CP$ violation is  difficult.

  Although we have calculated  $\Delta P$ numerically, 
 the approximate formula
  is useful to investigate the qualitative structure of the matter effect.
  The formulae have been given in the lowest order approximation 
  by Arafune, Koike and Sato \cite{AS}.
  In the case of $aL/2E\ll 1$ and $\Delta m_{21}^2 L/2E\ll 1$ with
   $a, \ \Delta m_{21}^2 \ll \Delta m_{32}^2$,  
  we show two differences of  transition probabilities between 
 $CP$-conjugate channels as follows:
 \begin{eqnarray}  
 \Delta P(\n_e \Ar \n_\m)&\equiv& P(\n_e \Ar \n_\m)-P(\bar\n_e \Ar \bar\n_\m)
  \nonumber \\
  &\simeq& P_m s^2_{23} + \Delta P_{CP} \ , \nonumber\\
\Delta P(\n_\m \Ar \n_e)&\equiv& P(\n_\m \Ar \n_e) - P(\bar\n_\m \Ar \bar\n_e)
  \nonumber \\
  &\simeq& P_m s^2_{23} - \Delta P_{CP} \ , \nonumber\\
 \Delta P(\n_e \Ar \n_\tau)&\equiv& 
 P(\n_e \Ar \n_\tau)- P(\bar\n_e \Ar \bar\n_\tau) \nonumber \\
  &\simeq& P_m c^2_{23} - \Delta P_{CP}  \ , 
  \label{MatterCP}
 \end{eqnarray}  
 \noindent   where
 \begin{equation}
 P_m = (1-2 s_{13}^2) c_{13}^2 s_{13}^2 \left ( 
 \frac{16 a}{\Delta m_{31}^2}\sin^2 \frac{\Delta m_{31}^2L}{4E} -
 \frac{4aL}{2E}\sin \frac{\Delta m_{31}^2L}{2E} \right ) \ ,
 \end{equation}
 \noindent and 
 $\Delta P_{CP}$ is the genuine $CP$ violating effect $4 J^{\n}_{CPl} f_{CP}$.
Since the matter effect in $\Delta P(\n_\m \Ar \n_\tau)$ is different from
$P_m$, we do not discuss this process.
  Approximate equations in eqs.(\ref{MatterCP}) suggest 
   how to extract the genuine $CP$ violating effect from the experimental data.
  By use of these equations, we get 
  \begin{equation}
\Delta  P_{CP} = \frac{1}{2}
 \left [\Delta P(\n_e \Ar \n_\m)- \Delta P(\n_\m \Ar \n_e)\right ] \ , 
  \label{EqT}
 \end{equation}
  \begin{equation}
\Delta  P_{CP} = \frac{1}{2}
 \left [(\cos 2\theta_{23}+1)\Delta P(\n_e \Ar \n_\m)+
        (\cos 2\theta_{23}-1)\Delta P(\n_e \Ar \n_\tau)\right ] \ .
\label{EqCP}
 \end{equation}
 \noindent
 These formulae are very useful to exculde  the matter effect.
 If $s_{23}$ will be determined  precisely in LBL experiments,
  these formulae become a test of
 the $CP$ violating phase in the three family model.
 The eq.(\ref{EqT}) is guaranteed exactly by the T violation relation
 as far as we use constant matter density.
Since  eq.(\ref{EqCP}) is an approximate formula
 with  $a\ll \Delta m_{31}^2$ in principle,
 we have to test that these are  well satisfied in  relevant parameter 
 regions without any approximation.
 In fig.4, we show the expected 
  $\Delta P(\n_e \Ar \n_\m)$ and $\Delta P(\n_e\Ar \n_\tau)$
  as well as $4 J^{\n}_{CP} f_{CP}$ in the case of $s_{12}=0.5$, 
 $s_{23}=1/\sqrt{2}$
  and $\phi=-90^{\circ}$. 
  The difference of 
  $|\Delta P(\n_e \Ar \n_\m)|$ and $|\Delta P(\n_e\Ar \n_\tau)|$ is 
 due to the matter effect
  because  $\Delta P(\n_e \Ar \n_\m)=-\Delta P(\n_e\Ar \n_\tau)$ is expected 
in vacuum.
  This difference is advantage to getting the genuine $CP$ violation.
 In fig.5, we show  $\Delta P_{CP}$ and the $4 J^{\n}_{CP} f_{CP}$, in which 
 $\Delta P_{CP}$ is estimated by using  eq.(\ref{EqCP}).  
 The $\Delta P(\n_e \Ar \n_\m)$ and
 $\Delta P(\n_e \Ar \n_\tau)$ are calculated numerically without any 
 approximation.
 As seen in fig.5, the calculated $\Delta P_{CP}$ agrees to 
$4 J^{\n}_{CP} f_{CP}$  within $1\%$.

Thus, eqs.(\ref{EqT}) and (\ref{EqCP}) provide a good test for 
    $CP$ violation in the three family model.
If we find the disagreement between
 $\Delta  P_{CP}$'s  obtained from both equations,
we should consider it as an evidence of  new physics.
 For example, our formulae are  modified
 if the sterile neutrino mixes with active neutrinos.
 The detail discussions will be presented elsewhere.

 %%%%%%%%%%%%%%%%%%%%%%%%%%%%%%%%%%%%%%%%%%%%%%%%%%%
 %%%%%%%%%%%%%%%%  Conclusion  %%%%%%%%%%%%%%%%%%%%%
 %%%%%%%%%%%%%%%%%%%%%%%%%%%%%%%%%%%%%%%%%%%%%%%%%%%
 We have studied the possibility to observe the genuine $CP$ violation 
 in a neutrino factory.
   Both $\Delta P(\n_\a \Ar \n_\b)$ and  $CP$ odd asymmetry $A_{CP}$
   have been  estimated  
  by taking account of the  large mixing angle MSW  solution.  
  The magnitude of $\Delta P(\n_e \Ar \n_\m)$ could be  $1\%$.
        If $s_{13}$ is lower than $0.05$, the matter effect is negligible
         in $\Delta P(\n_e \Ar \n_\m)$.
   If $A_{CP}$ is larger than $20\%$, the observation of 
        $P(\n_e \Ar \n_\m)$, which is smaller than $10^{-3}$, is  difficult.
        We have proposed how to extract the genuine $CP$ violation effect
         from  the neutrino oscillation data $\Delta P(\n_e \Ar \n_\m)$,
 $\Delta P(\n_\m \Ar \n_e)$ and $\Delta P(\n_e \Ar \n_\tau)$.
 The proposed method is expected to be advantage to observing $CP$ violation 
 in the neutrino factory.

\vskip 1 cm
 \section*{Acknowledgments}
I would like to thank E. Akhmedov for the useful comment on
the T violation.  I also thank J. Maalampi for useful discussions
 and his hospitality in Universty of Helsinki.
 This research is  supported by the Grant-in-Aid for Science Research,
 Ministry of Education, Science and Culture, Japan(No.10640274 ).
%%%%%%%%%%%%%%%%%%%%%%%%%%%%%%%%%%%%%%%%%%%%%%%%%%%%%%%%%%%%%%%%%
%%%%%%%%%%%%%%%%%%%%%%%%%%%%%%%%%%%%%%%%%%%%%%%%%%%%%%%%%%%%%%%%%
\newpage
%%%%%%%%%%

\newpage

%%%%%%%%%%%%%%%%%%%%%%%%%%%%%%%%%%%%%%%%%%%%
%%%%%%%%%%%%%%    Table 1   %%%%%%%%%%%%%%%%
%%%%%%%%%%%%%%%%%%%%%%%%%%%%%%%%%%%%%%%%%%%%
\begin{table}
\vskip 3 cm
\hskip -0.5  cm
\begin{tabular}{ | r| r| r| r| r| r| r| r| r| } \hline
          &   &  &      &        &         &        &        &           \\
$s_{13}\ $  & $P[{\rm vac}]\ \ $ & $\bar P[{\rm vac}]\ \ $ &
  $\Delta P[{\rm vac}]$ & 
$P[{\rm mat}]\ $ & $\bar P[{\rm mat}]\ $ &  $\Delta P[{\rm mat}]$ &
 $A_{CP}[{\rm vac}]$  &  $A_{CP}[{\rm mat}]$ \\
 & & & & & & & &      \\ \hline
      &         &       &        &        &          &        &     &    \\
$0.10$    & $0.00352$ & $0.00282$  & $0.00070$  & $0.00365$ & 
$0.00263$ & $0.00102$ & $0.11\quad $     & $0.16 \quad $       \\
     &         &       &        &        &          &        &     &    \\
$0.05$    & $0.00102$ & $0.00067$  & $0.00035$  & $0.00106$ & 
$0.00063$ & $0.00043$ & $0.21\quad $     & $0.26 \quad $       \\
     &         &       &        &        &          &        &     &    \\
$0.01$    & $0.00013$ & $0.00006$  & $0.000071$  & $0.00013$ & 
$0.00006$ & $0.000074$ & $0.36\quad $     & $0.38 \quad $       \\
  &      &   &   &    &      &       &     &       \\  \hline
\end{tabular}
\caption{Expected probabilities and asymmetries for the  $\n_e \Ar \n_\m$ 
oscillation  in the case 
 $s_{12}=0.5$, $s_{23}^2=0.5$, $\phi=-90^{\circ}$, 
$\Delta m_{21}^2=10^{-4}\eV^2$,   $\Delta m_{32}^2=3\times 10^{-3}\eV^2$ 
  with $E=7\G$ and $L=732 {\rm Km}$. 
 $\bar P$ denotes the $CP$-conjugate process.
$[{\rm vac}]$ and $[{\rm mat}]$ mean ``in the vacuum'' and 
``in the matter'', respectively.}
\end{table}
%%%%%%%%%%%%%%%%%%
\newpage
%%%%%%%%%%%%%%%  Figure 1  %%%%%%%%%%%%%%%%%
\begin{figure}
 \epsfxsize=12 cm
 \centerline{\epsfbox{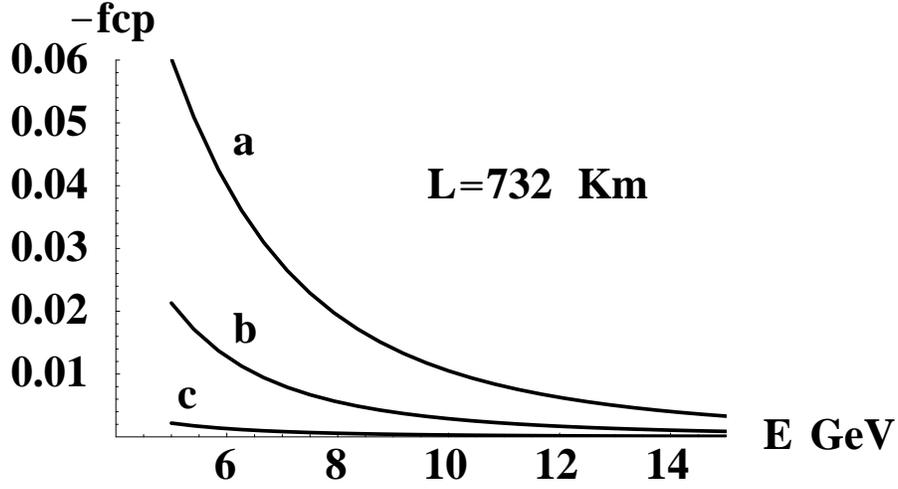}}
 \caption{Neutrino energy dependence of $f_{CP}$ in the cases of 
 {\bf a:}  $\Delta m_{32}^2=6\times 10^{-3}\eV^2$, 
 $\Delta m_{21}^2=10^{-4}\eV^2$,
 {\bf b:}  $\Delta m_{32}^2=3\times 10^{-3}\eV^2$, 
 $\Delta m_{21}^2=10^{-4}\eV^2$, 
 and {\bf c:}  $\Delta m_{32}^2=3\times 10^{-3}\eV^2$, 
$\Delta m_{21}^2=10^{-5}\eV^2$.}
\end{figure}
%%%%%%%%%%%%%%%%%%%%%%%%%%%%%%%%%%%%%%%%%%%%
%%%%%%%%%%%%%%%  Figure 2  %%%%%%%%%%%%%%%%%
\begin{figure}
 \epsfxsize=12 cm
 \centerline{\epsfbox{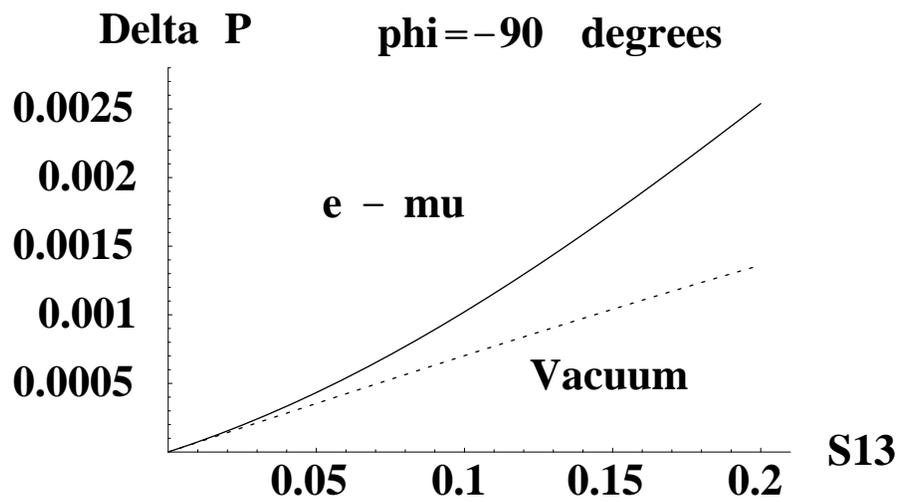}}
\caption{The $s_{13}$ dependence of  $\Delta P(\n_e \Ar \n_\m)$.
  The dashed-curve denotes the vacuum oscillation.}
\end{figure}
%%%%%%%%%%%%%%%%%%%%%%%%%%%%%%%%%%%%%%%%%%%%
%%%%%%%%%%%%%%%  Figure 3  %%%%%%%%%%%%%%%%%
\begin{figure}
 \epsfxsize=12 cm
 \centerline{\epsfbox{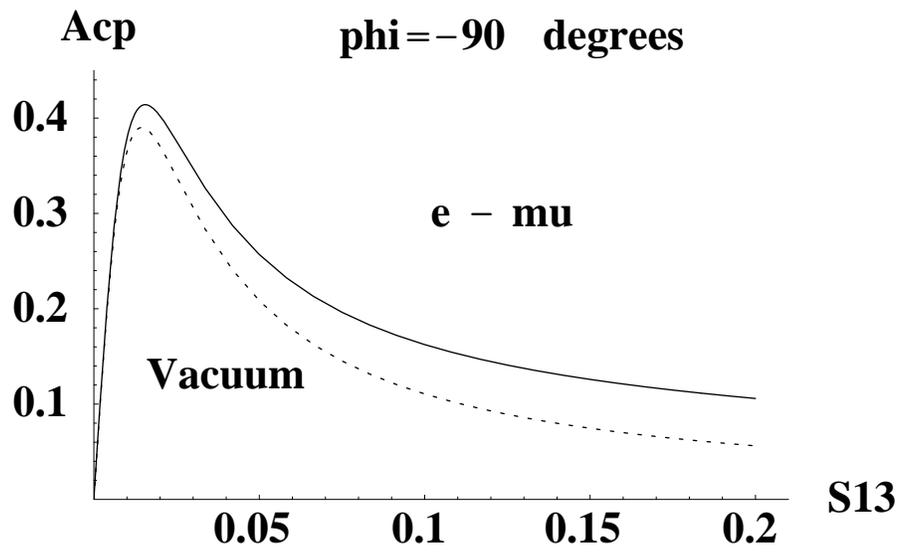}}
  \caption{The asymmetry of $\n_e \Ar \n_\m$ and $\bar\n_e \Ar \bar\n_\m$.
 The dashed-curve denotes the vacuum oscillation.}
 \end{figure}
%%%%%%%%%%%%%%%%%%%%%%%%%%%%%%%%%%%%%%%%%%%%
%%%%%%%%%%%%%%%  Figure 4  %%%%%%%%%%%%%%%%%
\begin{figure}
 \epsfxsize=12 cm
 \centerline{\epsfbox{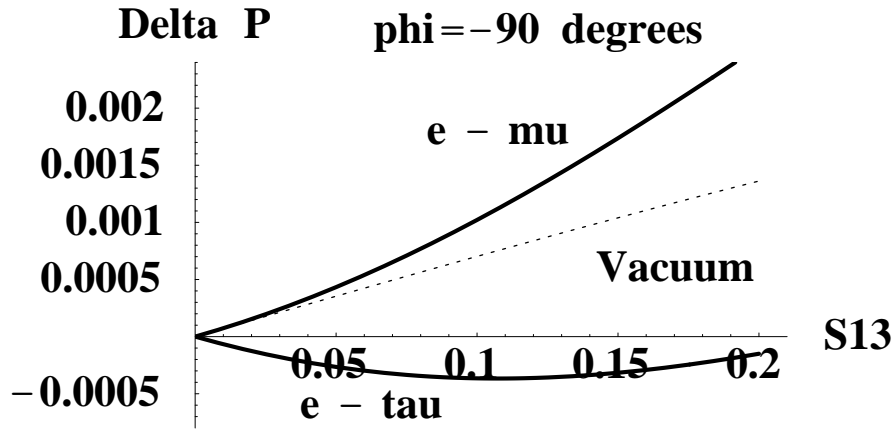}}
 \caption{The $s_{13}$ dependences of $\Delta P(\n_e \Ar \n_\m)$ and
 $\Delta P(\n_e \Ar \n_\tau)$ including the matter effect.  
Parameters are fixed as
  $s_{12}=0.5$,  $s_{23}^2=0.5$,  $\Delta m_{21}^2=10^{-4}\eV^2$,
 $\Delta m_{32}^2=3\times 10^{-3}\eV^2$ with   
 $E=7\G$  and  $L=732 {\rm Km}$.
  The dashed-curve denotes the vacuum oscillation.}
\end{figure}
%%%%%%%%%%%%%%%%%%%%%%%%%%%%%%%%%%%%%%%%%%%%
%%%%%%%%%%%%%%%  Figure 5  %%%%%%%%%%%%%%%%%
\begin{figure}
 \epsfxsize=12 cm
 \centerline{\epsfbox{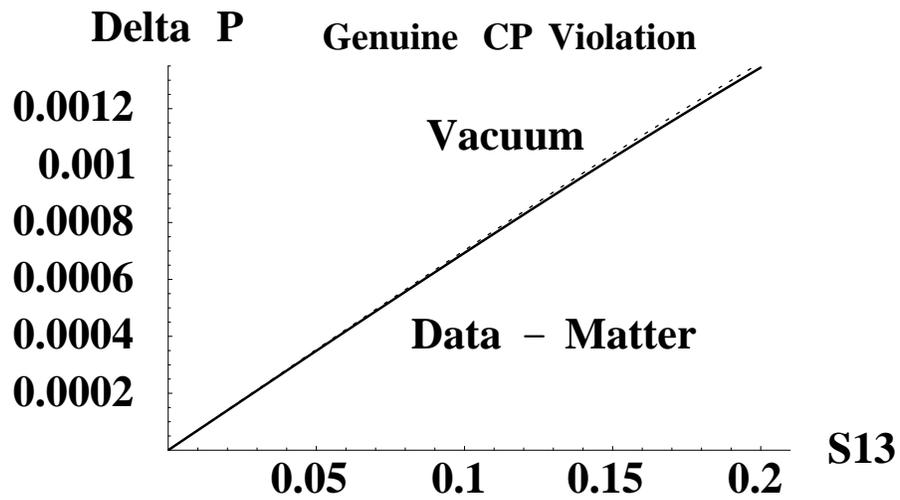}}
 \caption{The estimated genuine $CP$ violation,  which is given 
 by subtracting the matter effect from the data by using eq.(12).}
 \end{figure}
%%%%%%%%%%%%%%%%%%%%%%%%%%%%%%%%%%%%%%%%%%%%
\end{document}